\documentclass[%
 twocolumn,
superscriptaddress,
 amsmath,amssymb,
 aps, 
 physrev,
reprint,
]{revtex4-2}

\usepackage{mathrsfs}
\usepackage{graphicx}
\usepackage{dcolumn}
\usepackage{bm}

\usepackage{xcolor}
\usepackage{amssymb}
\usepackage[normalem]{ulem}
\usepackage{cancel}

\begin{document}

\preprint{APS/123-QED}

\title{\textbf{Training single-electron and single-photon stochastic physical neural networks}
} 

\author{T.~Dou}
\email{tong.dou@unsw.edu.au}
\affiliation{School of Engineering and Technology, UNSW Canberra, Campbell, ACT 2600, Australia}

\author{S.~Kumara}
\affiliation{School of Engineering and Technology, UNSW Canberra, Campbell, ACT 2600, Australia}

\author{J.~Burns}
\affiliation{Sussex Centre for Quantum Technologies, University of Sussex, Brighton BN1 9RH, United Kingdom}

\author{E.~Sigler}
\affiliation{School of Engineering and Technology, UNSW Canberra, Campbell, ACT 2600, Australia}

\author{P.~Girdhar}
\affiliation{School of Engineering and Technology, UNSW Canberra, Campbell, ACT 2600, Australia}
\affiliation{Department of Engineering Science, University of Oxford, Oxford OX1 3PJ, United Kingdom}

\author{D.~Petty}
\affiliation{School of Engineering and Technology, UNSW Canberra, Campbell, ACT 2600, Australia}

\author{G.~J.~Milburn}
\affiliation{Sussex Centre for Quantum Technologies, University of Sussex, Brighton BN1 9RH, United Kingdom}
\affiliation{National Quantum Computing Centre, Rutherford Appleton Laboratory, Didcot OX11 0QX, United Kingdom}

\author{J.~Plested}
\affiliation{School of Systems and Computing, UNSW Canberra, Campbell, ACT 2600, Australia}

\author{M.~J.~Woolley}
\email{m.woolley@unsw.edu.au}
\affiliation{School of Engineering and Technology, UNSW Canberra, Campbell, ACT 2600, Australia}


\date{\today}

\begin{abstract}
The computational demands of deep learning motivate the investigation of alternative approaches to computation.
One alternative is physical neural networks~(PNNs), in which learning and inference are performed directly via physical processes.
Stochastic PNNs arise when the underlying neurons are realized by the dynamics of a stochastic activation switch.
Here we propose novel electronic and photonic stochastic neurons.
The electronic realization is implemented by single-electron tunneling through a quantum dot.
The photonic realization is implemented via a single-photon source driving one of two modes coupled via a controllable beam-splitter-like interaction.
In the electronic case, the charge state of the quantum dot forms the basis for the stochastic neuron, whereas in the photonic case the occupation of the undriven mode serves as the basis for the stochastic neuron.
Training of stochastic PNNs is performed with models of stochastic neurons, as well as with coherently-driven, single-photon detector stochastic neurons previously introduced.
Several training strategies for MNIST handwritten digit classification have been investigated using single-hidden-layer stochastic PNNs, including varying the number of trials in each layer to control forward pass stochasticity and employing either true probability or empirical outputs in the backward pass to evaluate their influence on gradient estimation.
We show that when empirical outputs are used in the backward pass, the network achieves more than 97\% test accuracy with few trials per layer.
Despite the simplicity of the model architecture, high test accuracy is maintained in the presence of a high degree of noise and model uncertainty.
The results demonstrate the potential of embracing stochastic PNNs for deep learning.
\end{abstract}

\maketitle


\section{Introduction}
Deep neural networks have become a dominant paradigm for perception, inference, and control, driven by advances in data availability, algorithmic innovations, and specialized digital accelerators~\cite{lecun2015deep,schmidhuber2015deep,goodfellow2016deep}.
At the same time, the rising computational and energetic costs of training and inference~\cite{patterson2021carbon} motivate the exploration of alternative computing architectures beyond conventional digital computation.
One appealing route is to perform parts of neural network computation directly, using the natural physical dynamics of an engineered system.
These so-called physical neural networks (PNNs)~\cite{wetzstein2020inference,wright2022deep,momeni2023backpropagation,kalinin2025analog} enable potentially energy-efficient and ultrafast physical computation.

A central challenge with PNNs is training in the presence of device nonidealities~\cite{momeni2025training}.
Real hardware is noisy, lossy, and imperfectly calibrated; moreover, its characteristics drift in time and vary across nominally identical devices.
In many analog processors, these effects are often treated as small perturbations to an underlying deterministic computation, enabling ``noise-aware'' or ``quantization-aware'' variants of backpropagation~\cite{jacob2018quantization,wright2022deep,wang2022optical,zheng2023dual,wang2025asymmetrical,xu2026perfecting}.
This perturbative viewpoint can be effective when the signal-to-noise ratio (SNR) is high in every layer, so that randomness merely blurs otherwise deterministic activations.

However, when the operating point is pushed toward extreme energy efficiency or when the underlying information carriers are intrinsically discrete, the perturbative picture breaks down.
In optics, shot noise emerges from the quantization of the electromagnetic field, while in electronics, charge discreteness and stochastic tunneling are fundamental.
In these regimes the neuron output may become highly stochastic, with low per-evaluation SNR and discrete readout outcomes.
Rather than viewing this as an impairment, one can view the hardware naturally as implementing a stochastic neural network~\cite{bengio2013estimating,hubara2016binarized,milburn2022physics,ma2025quantum}, whose stochastic neurons are probabilistic activation switches.
The practical question then becomes: can we reliably train PNNs in a regime where stochasticity is the rule, not the exception?

Recent work on few-quanta optical neural networks has shown that high-accuracy, deterministic inference can be achieved, even when intermediate layers operate at SNR $\sim 1$, by explicitly incorporating a physics-based probabilistic model of the measurement process during training~\cite{ma2025quantum}.
This \emph{physics-aware} stochastic training perspective is compelling because it aligns the learning algorithm with the actual statistics of the experimental hardware.
However, in many physical platforms the training loop faces a different bottleneck: the experimentally accessible observables may be limited to samples of stochastic outputs, rather than direct access to the underlying activation probabilities or pre-activations.
Consequently, the learner may not have access to a sufficiently accurate probabilistic model of the device response, especially when the number of samples is limited.
This raises a basic and practically consequential problem: to what extent can a stochastic PNN be trained from limited sampling data, especially when the output layer is at a low SNR?

Here we study training strategies for stochastic physical neural networks, focusing on the regime where each neuron evaluation yields a discrete, stochastic outcome.
We introduce physically motivated realizations of stochastic neurons:
single-electron transistor stochastic neurons, where the charge state of a quantum dot acts as the stochastic neuron;
stochastic neurons based on controllable beam-splitter-like interaction driven by a deterministic single-photon source; 
and coherently-driven single-photon detector stochastic neurons previously introduced.
They share a common interface: a controllable pre-activation parameter that determines an activation probability and a measurable stochastic output sample.

Following a description of these physical systems, we systematically compare backpropagation compatible estimators under experimentally realistic constraints.
The results show that useful learning signals can be extracted even when only a small number of samples is available per layer, provided that the estimator is matched to the accessible observables and the network is trained with an appropriate treatment of stochasticity across layers.
In addition, high accuracy can be maintained despite substantial noise and model uncertainty.

\section{Physical Stochastic Neurons}\label{PhyStoNeuron}
Artificial neural networks are commonly formulated as compositions of affine transformations followed by nonlinear activation functions~\cite{goodfellow2016deep}.
For a fully-connected feedforward neural network, the forward propagation at the $l^{\text{th}}$ layer can be written as
\begin{subequations}
\begin{align}
    \mathbf{z}^{(l)} &= \mathbf{W}^{(l)}\mathbf{h}^{(l-1)}+\mathbf{w}_{0}^{(l)}, \\
    \mathbf{h}^{(l)} &= \bm{f}\left(\mathbf{z}^{(l)}\right),
\end{align}
\end{subequations}
where $\mathbf{z}^{(l)}$ is the pre-activation, $\mathbf{W}^{(l)}$ is the weight matrix, $\mathbf{w}_{0}^{(l)}$ is the bias vector, $\mathbf{h}^{(l-1)}$ is the activation from the previous layer, and $\bm{f}(\cdot)$ is an element-wise nonlinear activation function.
The overall transformation implemented by a deep neural network is a nested composition of such layer-wise maps.

In PNNs, these mathematical transformations can be implemented by physical processes directly realized in hardware.
In so-called isomorphic PNNs~\cite{wright2022deep}, the network topology and the set of tunable physical parameters are designed to mirror the connectivity and parameters of the corresponding digital neural network.
A key distinction between digital and physical modules is that physical systems are unavoidably subject to noise.
This leads to a mismatch between the mathematical and physical transformations, even when the same nominal parameters are assigned.
While certain noise sources can be mitigated through engineering improvements, others are fundamentally linked to the discrete nature of the information carriers themselves.
For example, in optical systems, this discreteness manifests as photon shot noise, whereas in electronic systems it arises from the quantization of charge.
When operating neural-network hardware at extremely low energy levels, which is a desirable regime for scalable and energy-efficient computing, such shot noise becomes a dominant effect rather than a small perturbation.
As a result, the output of a physical neuron is often stochastic, even when the input is held fixed.

These considerations motivate the concept of a stochastic PNN, in which neurons are implemented as physical stochastic, rather than deterministic, nonlinear elements~\cite{ma2025quantum}.
We refer to the neurons in such a network as physical stochastic neurons~(PSNs).
In this setting, stochasticity is not merely tolerated but becomes an integral part of the neuron.
Specifically, in many physically relevant cases, the activation of a PSN produces a binary output $h_{i}\in\{0,1\}$, with the probability of activation given by a function of the pre-activation $z_{i}$ evaluated at the $i^{\text{th}}$ neuron in the layer.
That is,
\begin{equation}\label{eq:PSN_output}
\begin{split}
h_{i} = f_{\text{PSN}}(z_{i}) &= \text{Bernoulli}(p_{\text{PSN}}(z_{i}))
\\[4pt]
&
=
\begin{cases}
    1, & \text{with probability } p_{\text{PSN}}(z_{i}), \\
    0, & \text{otherwise},
\end{cases}
\end{split}
\end{equation}
where $p_{\text{PSN}}\left(z_{i}\right)$ is a function mapping the pre-activation to a probability in the interval $[0,1]$.
It may take different forms depending on the underlying physical mechanism.
Within this framework, the stochastic activation can be interpreted as a probabilistic switch, also known as an activation switch.

Next we consider three specific implementations of PSNs: the single-photon detector neuron~\cite{ma2025quantum}, the single-electron transistor neuron~\cite{milburn2022physics}, and introduce what we call the true single-photon neuron, which is based on a deterministic single-photon source.
Each of these neurons implements a stochastic activation function, which maps the pre-activation $z_{i}$ to a binary output $h_{i} \in \{0,1\}$ according to some probability distribution $p_{\text{PSN}}(z_{i})$.

\subsection{Single-photon detector stochastic neuron}
\begin{figure}[!htbp]
\includegraphics[width=0.6\linewidth]{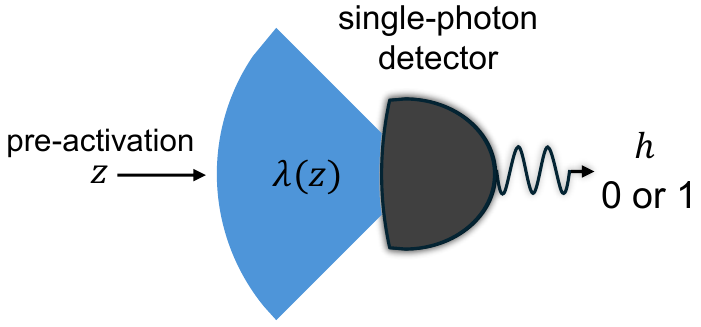}
\caption{\label{fig:SPD_neuron} 
Schematic of an SPD stochastic neuron adapted from~\cite{ma2025quantum}.
The mean photon number $\lambda(z)$ incident on the detector in a given measurement interval is a function of the pre-activation $z$.
The output of the SPD neuron is modeled as the stochastic output from the single-photon detector, given by $h=f_{\text{SPD}}(z)$, where $h=0$ corresponds to no-clock and $h=1$ corresponds to a click.}
\end{figure}

Single-photon detector (SPD) stochastic neurons have been introduced and demonstrated in photonic neural networks operating at the few-photon level~\cite{ma2025quantum}. 
A schematic of the SPD neuron is shown in Fig.~\ref{fig:SPD_neuron}.
Here, it is adopted as a reference for comparison with other PSNs.

In SPD stochastic neurons, the stochasticity originates from the process of photon counting.
For coherent light, photon counting is a Poisson process~\cite{WallsMilburnQuantumOptics3e}.
Specifically, under coherent illumination, the probability that a single-photon detector detects $n$ photons within a given measurement window is given by
\begin{equation}
P(n) = \frac{\lambda^{n}e^{-\lambda}}{n!},
\end{equation}
where $\lambda$ denotes the mean photon number incident on the detector within a measurement window.
The probability of detecting no photons is therefore $P(0)=e^{-\lambda}$, corresponding to a no-click event. 
Therefore, the probability of detecting one or more photons, corresponding to a click event, is
\begin{equation}
P_{\mathrm{click}} = 1-P(0) = 1 - e^{-\lambda} = p_{\mathrm{SPD}}(\lambda).
\end{equation}

In this click-detector setting, the photon detection process can be modeled as sampling of the click probability $P_{\mathrm{click}}(\lambda)$.
For disjoint measurement windows, the corresponding trials are independent, which enables repeated independent sampling (Bernoulli trials) of the stochastic neuron output.
The output of an SPD stochastic neuron in a single measurement window can be described as
\begin{equation}
    f_{\mathrm{SPD}}(z_{i}) =
    \begin{cases}
    1, & \text{with probability } p_{\mathrm{SPD}}(\lambda(z_{i})), \\
    0, & \text{otherwise},
    \end{cases}
\end{equation}
where the function $\lambda(z_{i})$ maps the pre-activation value $z_{i}$ at the $i^{\text{th}}$ neuron, obtained from the preceding linear transformation, to a mean photon number incident on the detector in a given time interval.
In a single-shot measurement, an SPD produces a binary outcome $h_{i}$ corresponding to the presence or absence of a detection click.

The specific form of $\lambda(z_{i})$ depends on the optical encoding scheme employed.
Two setups are considered in~\cite{ma2025quantum}: an incoherent setup where the pre-activation is encoded in the intensity of light, leading to $\lambda(z_{i})=z_{i}$, and a coherent optical setup where the pre-activation is encoded in the amplitude of light while the intensity is detected, yielding $\lambda(z_{i})=|z_{i}|^{2}$.
Here we adopt the coherent implementation, as incoherent intensity encoding imposes non-negativity constraints on the weights, whereas coherent encoding relaxes this restriction and leads to better performance.

\subsection{Single-electron transistor stochastic neuron}
Here we introduce  a single-electron transistor (SET)~\cite{ferry2009transport} stochastic neuron as an electronic realization of stochastic neurons operating in the single-electron regime~\cite{darulova2020autonomous,milburn2022physics}.
In such devices, stochasticity arises from the discrete nature of electric charge and the probabilistic nature of single-electron tunneling events.

\begin{figure}[!htbp]
\includegraphics[width=0.8\linewidth]{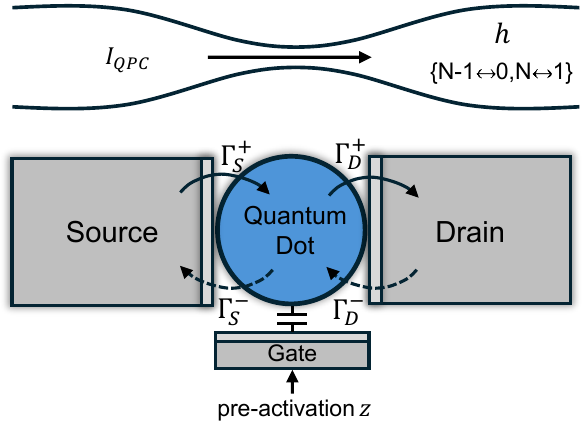}
\caption{\label{fig:SET_neuron} Schematic of a SET stochastic neuron realized in a semiconductor quantum dot system. 
The pre-activation $z$ modulates the dot energy level $\varepsilon(z)$ via the gate electrode. 
Electron tunneling between the dot and leads (source and drain) is governed by the tunneling rates $\Gamma_{\text{S/D}}^{\pm}$, where the total electron tunneling rates in and out of the dot are defined as $\Gamma^{\text{in}}=\Gamma_{\text{S}}^{+}+\Gamma_{\text{D}}^{-}$ and $\Gamma^{\text{out}}=\Gamma_{\text{D}}^{+}+\Gamma_{\text{S}}^{-}$, respectively.
The resulting dot occupation can be modeled as the output of the SET stochastic neuron, which is monitored by the current $I_{\text{QPC}}$ of a nearby quantum point contact~(QPC), provided by $h=f_{\text{SET}}(z)$, where $h=0$ denotes the unoccupied state ($N-1$ electrons) and $h=1$ denotes the occupied state ($N$ electrons) of the quantum dot.
}
\end{figure}

We consider a semiconductor quantum dot operated in the Coulomb blockade regime, where the charging energy of the dot dominates both thermal fluctuations and the source-drain voltage bias.
Under the constant-interaction model and assuming fast internal relaxation, the dot maintains a well-defined electron number, allowing at most one excess electron to occupy the dot at any given time at an energy level $\varepsilon$.
The level can be modulated via electrode voltages and thus facilitates encoding of the pre-activation via $\varepsilon(z)$.
In the weak tunnel-coupling limit, electron transport occurs via incoherent tunneling (i.e., sequential tunneling) between the dot and source/drain, and the dot dynamics can be described as a random telegraph process corresponding to unoccupied and occupied states.

A schematic of the quantum-dot based SET stochastic neuron is given in Fig.~\ref{fig:SET_neuron}.
Let $P_0$ and $P_1$ denote the probabilities that the dot is unoccupied or occupied, respectively.
The stochastic dynamics of this two-state system can be described by a master equation,
\begin{equation}
\frac{d}{dt}
\begin{bmatrix}
P_0 \\
P_1
\end{bmatrix}
=
\begin{bmatrix}
-\Gamma^{\mathrm{in}} & \Gamma^{\mathrm{out}} \\
\Gamma^{\mathrm{in}} & -\Gamma^{\mathrm{out}}
\end{bmatrix}
\begin{bmatrix}
P_0 \\
P_1
\end{bmatrix},
\end{equation}
where $\Gamma^{\mathrm{in}}$ and $\Gamma^{\mathrm{out}}$ denote the tunneling rates for electrons in and out of the dot, respectively.

Assuming the leads act as Fermi reservoirs in thermal equilibrium, the tunneling rates across the source and drain barriers, as shown in Fig.~\ref{fig:SET_neuron}, are given by
\begin{equation}
\begin{aligned}
\Gamma_{\mathrm{S}}^{+} &= \Gamma_{\mathrm{S}}\, \left[1-n_{F}(\varepsilon)\right], 
&\quad
\Gamma_{\mathrm{D}}^{-} &= \Gamma_{\mathrm{D}}\, \left[1-n_{F}(\varepsilon)\right], \\
\Gamma_{\mathrm{S}}^{-} &= \Gamma_{\mathrm{S}}\,n_{F}(\varepsilon),
&\quad
\Gamma_{\mathrm{D}}^{+} &= \Gamma_{\mathrm{D}}\, n_{F}(\varepsilon),
\end{aligned}
\end{equation}
where $n_{F}(\varepsilon)=1/{\left(1+e^{\varepsilon/k_{B}T}\right)}$ is the Fermi--Dirac occupation of the dot energy level, $k_{B}$ is the Boltzmann constant, and $T$ is the temperature. 
Consequently, the total in and out rates of the dot are $\Gamma^{\text{in}} = \Gamma_{\text{S}}^{+}+\Gamma_{\text{D}}^{-}$ and $\Gamma^{\text{out}} = \Gamma_{\text{D}}^{+}+\Gamma_{\text{S}}^{-}$.
The total tunneling rates can thus be written as
\begin{equation}
    \Gamma^{\mathrm{in}} = \Gamma \, \left[1 - n_{F}(\varepsilon)\right] ,\quad \Gamma^{\mathrm{out}} = \Gamma \, n_{F}(\varepsilon),
\end{equation}
where $\Gamma = \Gamma_{\text{S}} + \Gamma_{\text{D}}$ is the total coupling strength to the source and drain leads.
In the steady-state, the probabilities of the single energy level of the dot being unoccupied and occupied are given by
\begin{subequations}
\label{dot-ss}
    \begin{align}
        P_{0}^{\mathrm{ss}} &= \frac{\Gamma^{\mathrm{out}}}{\Gamma^{\mathrm{in}} + \Gamma^{\mathrm{out}}} = n_{F}(\varepsilon) \\
        P_{1}^{\mathrm{ss}} &= \frac{\Gamma^{\mathrm{in}}}{\Gamma^{\mathrm{in}} + \Gamma^{\mathrm{out}}} = 1-n_{F}(\varepsilon)
    \end{align}
\end{subequations}
This result shows that the steady-state probabilities are governed by the Fermi function of the energy level $\varepsilon$.

Using the symmetry property of the Fermi-Dirac distribution, $1-n_{F}(\varepsilon)=n_{F}(-\varepsilon)$, $P_1^{\mathrm{ss}}$ can be expressed in terms of the sigmoid function as
\begin{equation} 
P_1^{\mathrm{ss}}(\varepsilon) = \sigma\left( \frac{\varepsilon }{k_{B} T} \right).
\end{equation}

Identifying the dot energy $\varepsilon$ as a function of the pre-activation $z_{i}$, the output of a SET stochastic neuron in a single measurement can be written as
\begin{equation}
f_{\mathrm{SET}}(z_{i})
=
\begin{cases}
1, & \text{with probability }  p_{\mathrm{SET}}(z_{i}), \\
0, & \text{otherwise},
\end{cases}
\end{equation}
where $p_{\mathrm{SET}}(z_{i})=P_1^{\mathrm{ss}}(\varepsilon(z_{i}))$. 
Without loss of generality, we use the sigmoid function to encode the activation probability of the SET stochastic neuron with $\varepsilon(z_{i}) = k_{B}T z_{i}$.

It is assumed here that individual electron tunnel events can be resolved by the read-out apparatus.
The output of the SET neuron can be measured, for example, via a nearby quantum point contact~(QPC), whose current $I_{\text{QPC}}$ is sensitive to the dot's occupation.
The $I_{\text{QPC}}$ switches between two levels, corresponding to the dot's occupation state $h_{i}\in\{0,1\}$.

\subsection{True single-photon stochastic neuron}\label{sec:TSP_neuron}
Next we introduce a true single-photon~(TSP) stochastic neuron, built upon a deterministic single-photon source~\cite{loredo2026deterministic,SPS} and a controllable, beam-splitter-like interaction Hamiltonian between two bosonic modes $a$ and $b$~\cite{BM-switch}.
The stochasticity arises from measurement of the $b$ mode occupation.

The interaction Hamiltonian describing this system is given by
\begin{equation}\label{eq:Hamiltonian}
    H=\alpha(t) a b^\dagger + \alpha^*(t) a^\dagger b,
\end{equation}
 where $a$ is the annihilation operator for a photonic mode, $b$ is the annihilation operator for a generalised bosonic mode, and $\alpha(t)$ represents a time-dependent coupling strength.
 This interaction Hamiltonian can be realized in a number of technologies, including an electro-optic beam-splitter~\cite{psiquantum2025manufacturable} and coupled superconducting circuits~\cite{lu2023high}.  
 
 Another possible realization, an optomechanical system~\cite{Sonar:25,Zivari:25}, is depicted schematically in Fig.~\ref{fig:TSP_neuron}.
 Here $a$ refers to photons in an optical cavity mode while $b$ refers to phonons in a mechanical oscillator. 
 The system is driven by an incident single photon field coupled to the mode $a$, whilst both modes $a$ and $b$ are irreversibly coupled to decay channels with decay rates $\kappa$ and  $\gamma$, respectively.  
 Now the control field, described by $\alpha(t)$, takes the form of a coherent pulse co-propagating with the single photon pulse.  

To analyze this neuron, we have to incorporate a description of the single-photon field.
 {A single, continuous-mode photon state is defined as a coherent superposition of a single excitation superposed over many frequency modes~\cite{LoudonQuantumTheoryOfLight3e}
 \begin{equation}\label{eq:single_photon_state}
     |1_\xi\rangle=\int d\omega \tilde{\xi}(\omega) a_\omega^\dagger|0\rangle.
 \end{equation}
 The average field of such a state is zero but the probability to detect the photon per unit time is proportional to $|\xi(t)|^2$ where $\xi(t)$ is the inverse Fourier transform of $\tilde{\xi}(\omega)$. 
 The input was chosen to be an exponentially-decaying single-photon field with temporal decay rate $\zeta$ such that
\begin{equation}
\xi(t)=
    \begin{cases}
        \sqrt{\zeta}e^{-\zeta t/2},&t\geq 0;\\
        0,&t<0.
    \end{cases}
\end{equation}

 \begin{figure}[!htbp]
\includegraphics[width=0.8\linewidth]{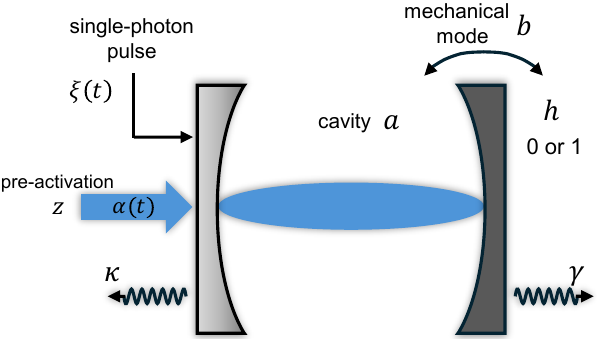}
\caption{\label{fig:TSP_neuron} Schematic of a TSP stochastic neuron realized in an optomechanical system.
The neuron is driven by an incident single-photon pulse $\xi(t)$ with a temporal decay rate $\zeta$. The pre-activation $z$ modulates the interaction between the cavity mode $a$ and the mechanical mode $b$ via the time-dependent coherent control pulse $\alpha(t)$.
The system evolution is subject to the cavity decay rate $\kappa$ and mechanical dissipation $\gamma$.
Driven by the  control pulse, the single-photon state is stochastically transferred to the mechanical mode, whose occupation at any time $t$ is at most unity, providing a stochastic binary neuron output.
The neuron activation can be obtained by measuring the mechanical occupation number, provided by $h=f_{\text{TSP}}(z)$, where $h=0$ denotes the unoccupied state and $h=1$ denotes the occupied state of mode $b$.
}
\end{figure}

To analyze the dynamics of the TSP neuron, we use the Fock state master equation (FSME) approach~\cite{gough2012quantum,Combes_2017}.
This approach tracks excitations of the external fields, the localized system modes $(a,b)$, and the coherences between them. 
We treat this as a scattering problem between the input state and output state,
$|1_\xi\rangle\otimes|0\rangle\rightarrow |\Phi(t)\rangle\otimes|0\rangle+|0\rangle\otimes|\Psi(t)\rangle$.
The output state is a superposition of two possible scenarios: either the photon remains in the environment while the local modes are in vacuum, or the environment is vacuum and the single excitation has been transferred to the local modes.

The evolution of some system operator $X$ at time $t$ is given by
\begin{equation}
    j_t(X)=U^\dagger(t)(X\otimes \mathbb{I}_{f})U(t),
\end{equation}
where $U(t)$ is the joint unitary time-evolution operator of the system modes and fields, and $\mathbb{I}_{f}$ is the identity operator acting on the environment modes.
Then the expectation values used in the FSME are defined as:
\begin{equation}
    \varpi^{ij}_t(X)=\langle \eta~\phi_{i}|j_t(X)|\eta~\phi_{j}\rangle,
\end{equation}
where $|\eta \rangle$ is the initial state of the system modes $(a,b)$, $i,j\in\{0,1\}$, and 
\begin{equation}\label{eq:varphi_as_expectation_1}
    |\phi_k\rangle = \begin{cases}
        |0\rangle,&k=0;\\
        |1_\xi\rangle,&k=1.
    \end{cases}
\end{equation} 

In the FSME approach, moments of system operators obey a hierarchy of equations given by
\begin{align}\label{eq:monent_omega}
    \frac{d}{dt} \varpi^{ij}\!\bigl(X(t)\bigr) &=\! \varpi^{ij}\!\bigl(\mathcal{L} X\bigr) +\sqrt{i}\,\xi^{*}(t)\,\varpi^{i-1,j}\!\bigl([X,\sqrt{\kappa}a]\bigr) \nonumber\\
    &\quad + \sqrt{j}\,\xi(t)\,\varpi^{i,j-1}\!\bigl([\sqrt{\kappa}a^\dagger,X]\bigr).
\end{align}
The indices $i,j\in\{0,1\}$ track the excitation, and $\mathcal{L}$ is the Lindbladian,
\begin{equation}\label{eq:lindbladian}
    \mathcal{L}X=i[X,H]+\mathcal{D}[\sqrt{\kappa}a]X + \mathcal{D}[\sqrt{\gamma}b]X,
\end{equation}
for dissipative superoperators $\mathcal{D}[C]X \equiv C^{\dagger} X C - \frac{1}{2}\left( C^{\dagger} C X + X C^{\dagger} C] \right)$ associated with dissipation channel $C$. 
These decay channels are mutually independent, so $\mathcal{D}[C]X=0$ if $X$ is independent of the mode associated with $C$. 
It is important to note that all decay channels in the system are included in the Lindbladian, whilst only the operator that couples to the single-photon drive are included in the commutator terms.

We need to calculate $\varpi^{11}_t(b^\dagger b)$, which is the time-dependent occupation of the $b$-mode $\langle b^\dagger b\rangle$.
It can be shown that
\begin{equation}
    \varpi^{11}_t(b^\dagger b)=\varpi^{10}_t(b^\dagger)^*\varpi^{10}_t(b^\dagger),
\end{equation}
which enables the computation of $\varpi^{11}_t(b^\dagger b)$ using the system described by Eqs.~\eqref{eq:monent_omega}--\eqref{eq:lindbladian}.
This factorization significantly simplifies the problem, as one only needs to solve for the off-diagonal element $\varpi^{10}_t(b^\dagger)$.

Solving the system of equations corresponding to Eqs.~\eqref{eq:monent_omega}--\eqref{eq:lindbladian} with an input exponential single-photon field and constant $\alpha(t)=\alpha$ yields, as shown explicitly in App.~\ref{append_ps_neuron}, the analytic expression for the $b$-mode occupation,
\begin{equation}
\label{eq:ps_activation}
\langle b^\dagger b\rangle \!=\! \frac{64\alpha^{2}\zeta \kappa}{\Delta^2}
\!\Bigg[
\frac{1 - e^{\frac{t}{4}\left(\Delta - u\right)}}{\Delta - u} 
+ 
\frac{1 - e^{\frac{t}{4}\left(\Delta + u \right)}}{\Delta + u}
\Bigg]^{2}
\!e^{-t \zeta},
\end{equation}
where we have introduced the notation
\begin{subequations}
    \begin{align}
        \Delta &= \sqrt{(\gamma - \kappa)^2 - 16\alpha^2},\\
        u &= \gamma  + \kappa -2\zeta.
    \end{align}
\end{subequations}

As there is only one excitation in the system, the expectation value $\langle b^{\dagger} b\rangle$ is strictly bounded between $0$ and $1$, representing the probability of finding a single excitation in the $b$-mode.
This allows the system to function as a stochastic neuron with activation probability,
\begin{equation}
    p_{\mathrm{TSP}}(z_{i}) = \langle b^\dagger b \rangle\left(\alpha(z_{i})\right).
\end{equation}
The pre-activation $z_{i}$ is encoded in the coupling strength $\alpha$ as $\alpha(z_{i})=z_{i}$.
Accordingly, the neuron activation can be modeled as,
\begin{equation}
    f_{\mathrm{TSP}}(z_{i}) = 
    \begin{cases}
        1, & \text{with probability } p_{\mathrm{TSP}}(z_{i}), \\
        0, & \text{otherwise}.
    \end{cases}
\end{equation}
Each $b$ mode occupation measurement corresponds to a sample of the neuron with a binary outcome $h_{i}$.

\section{Training Stochastic Physical Neural Networks}
While PSNs provide a natural interface between low-energy physical processes and neural networks, their intrinsically stochastic nature poses challenges for training.
In stochastic PNNs, the forward pass involves a probabilistic activation process.
To mathematically formulate it, given the pre-activation at the $l^{\text{th}}$ layer $\mathbf{z}^{(l)}$, the activation of the $i^{\text{th}}$ neuron can be modeled as
\begin{equation}
    h_{i}^{(l)} = f\left( p_{\text{PSN}} \left(z_{i}^{(l)}\right), u_{i}^{(l)} \right),
\end{equation}
where $u_{i}^{(l)}$ is a random variable representing the internal physical noise source.
Under standard backpropagation, the gradient of a loss function $L$ with respect to $\mathbf{z}^{(l)}$ is
\begin{align}\label{gz_PNN}
    \frac{\partial L}{\partial \mathbf{z}^{(l)}} &= \left(\frac{\partial \mathbf{h}^{(l)}}{\partial \mathbf{z}^{(l)}}\right)^{\top} \frac{\partial L}{\partial \mathbf{h}^{(l)}},
\end{align}
where $\mathbf{h}^{(l)}$ is sampled from $p_{\text{PSN}}\left(\mathbf{z}^{(l)}\right)$.
However, individual realizations of the activation $\mathbf{h}^{(l)}$ are discrete and non-differentiable, rendering standard backpropagation training inapplicable.

A strategy to address this challenge is to adopt ``physics-aware stochastic training''~\cite{ma2025quantum}.
If the pre-activation can be evaluated and the corresponding activation probability is accessible, one could bypass the sampling in the backward pass and instead propagate gradients through the expectation value of the stochastic neuron output.
For a Bernoulli PSN, the expectation value of the activation can be written as
\begin{align}
    \mathbb{E}\left[h_{i}^{(l)} \Big| z_{i}^{(l)}\right] &= \mathbb{E}\left[ f\left(p_{\text{PSN}}\left(z_{i}^{(l)}\right),u_{i}^{(l)}\right) \Big| z_{i}^{(l)} \right] \nonumber \\
    &= \int_{0}^{1} f\left( p_{\text{PSN}}\left(z_{i}^{(l)}\right), u_{i}^{(l)} \right) p(u_{i}^{(l)}) \mathrm{d}u_{i}^{(l)} \nonumber \\
    &= \int_{0}^{p_{\text{PSN}}\left(z_{i}^{(l)}\right)} 1 \mathrm{d}u_{i} + \int_{p_{\text{PSN}}\left(z_{i}^{(l)}\right)}^{1} 0 \mathrm{d}u_{i}^{(l)} \nonumber \\
    &= p_{\text{PSN}}\left(z_{i}^{(l)}\right),
\end{align}
where $u_{i}$ is assumed to be uniformly distributed in $[0,1]$.
This result shows that the expected output is a smooth and differentiable function of the pre-activation.
Based on this observation, one may replace the stochastic output by its expectation in the backward pass and compute
\begin{equation}\label{eq:TP_estimator}
    \frac{\partial L}{\partial \mathbf{z}^{(l)}} = \left(\frac{\partial p_{\text{PSN}}\left(\mathbf{z}^{(l)}\right)}{\partial \mathbf{z}^{(l)}}\right)^{\top} \frac{\partial L}{\partial \mathbf{h}^{(l)}}.
\end{equation}
This approach has been previously called \emph{deterministic mean-field} estimator in~\cite{ma2025quantum}, though we shall refer to it as the \emph{true probability}~(TP) approach.
Then, the corresponding gradients with respect to the weight $\mathbf{W}^{(l)}$ and bias $\mathbf{w}_{0}^{(l)}$ retain the same structure as in deterministic neural networks,
\begin{equation}\label{g_W_b}
    \frac{\partial L}{\partial \mathbf{W}^{(l)}} = \frac{\partial L}{\partial \mathbf{z}^{(l)}} \mathbf{h}^{(l-1) \top},\quad \frac{\partial L}{\partial \mathbf{w}_{0}^{(l)}} = \frac{\partial L}{\partial \mathbf{z}^{(l)}},
\end{equation}
where $\mathbf{h}^{(l-1)}$ is the activation from the previous layer.

However, in many physical systems, the activation probability $p_{\text{PSN}}$ is unknown, and only stochastic samples of the neuron output can be observed.
This motivates estimators of the gradient that operate under finite-sampling constraints.
After considering the TP approach, we introduce an \emph{empirical gradient}~(EG) estimator, which infers gradient information from sample statistics, and \emph{straight-through}~(ST) estimators, which bypasses both the stochastic sampling process and the activation nonlinearity in the backward pass by substituting a surrogate gradient.

Across all experiments in this study, we consider a stochastic PNN with a single hidden layer, trained on the MNIST handwritten digit dataset~\cite{lecun1998gradient}.
The model is implemented as a fully-connected feedforward neural network with a $784\text{-}400\text{-}10$ neuron architecture.
We implemented and trained our networks utilizing the PyTorch machine learning framework with a GPU.
Unless otherwise specified, the activation of the output layer is the softmax function, and the networks are trained using cross-entropy~(CE) loss.
We optimize the network using stochastic gradient descent with a learning rate of $0.001$ and a batch size of 128.

\subsection{True probability in backward pass}
\label{sec:benchmark}
We first establish a reference benchmark based on the TP approach.
During the forward pass, the intrinsically stochastic nature of the neurons is retained under finite-sampling constraints.
Specifically, for a fixed pre-activation of $z_{i}^{(l)}$, we can draw $K$ independent samples $\left\{h_{i,k}^{(l)}\right\}_{k=1}^K$ and construct the sample mean estimator of the neuron activation
\begin{equation}\label{eq:empirical_estimate}
    \hat{h}_{i}^{(l)}
    =
    \frac{1}{K}\sum_{k=1}^K h_{i,k}^{(l)}.
\end{equation}
This benchmark corresponds to the stochastic neuron activations being replaced by their expectation values during the backward passes.

The primary difference between the evaluated models lies in the form of the stochastic activation probability $p_{\text{PSN}}$ used in the hidden layer.
We compare the three PSNs introduced in Section~\ref{PhyStoNeuron}: the SPD, SET, and TSP stochastic neurons.
For benchmarking against the configuration in~\cite{ma2025quantum}, the output layer is evaluated in the infinite-trial limit.
Specifically, the pre-activations of the output layer are deterministically mapped to class probabilities using a softmax function.

\begin{figure}[!htbp]
\includegraphics[width=0.9\linewidth]{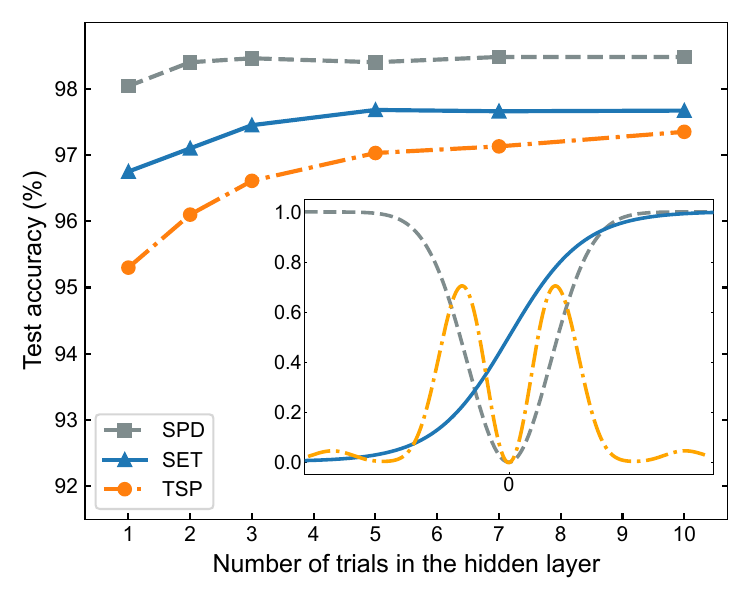}
\caption{\label{fig:benchmark} Benchmarking stochastic physics-aware training on SPD, SET, and TSP stochastic neurons. 
The TP approach is applied to both the hidden and output layers. 
Inset: Activation probabilities $p_{\text{PSN}}(z)$ as functions of the pre-activation $z$ for corresponding PSNs. 
The curves are plotted over different $z$-ranges: $[-3,3]$ for SPD, $[-5,5]$ for SET, $[-50,50]$ for TSP, and for TSP neuron, parameters are set to $t=0.21$, $\gamma=0.02$, $\kappa=30.0$, $\zeta=10.7$.
All three PSNs support effective training even with only 1 trial in the hidden layer, and test accuracy improves as the number of trials increases.}
\end{figure}

Fig.~\ref{fig:benchmark} shows the test accuracy using the TP approach for SPD, SET and TSP stochastic neurons, as a function of the number of trials used in the hidden layer.
All three activation probabilities support stable and effective training, and test accuracy consistently improves as the number of trials increases.
These results demonstrate that, in the presence of sampling noise, different PSNs can be trained using a physics-aware stochastic training approach, and that the specific form of the activation probability influences training efficiency.

\subsection{Empirical gradient estimator in backward pass}
\label{sec:EG}
As discussed, the TP approach provides an idealized reference for training stochastic PNNs by bypassing the sampling process while retaining the exact gradient structure of the activation probability.
However, in general, neither the activation probability $p_{\text{PSN}}(z)$ nor the pre-activation $z$ may be known.
Instead, only stochastic samples of the neuron output can be observed.
This motivates the development of gradient estimators that approximate the TP approach using finite samples.

Consider a stochastic PSN with output as in Eq.~\eqref{eq:PSN_output}.
In the TP approach, the derivative $\partial p_{\text{PSN}}\left(\mathbf{z}^{(l)}\right)/\partial \mathbf{z}^{(l)}$ in Eq.~\eqref{eq:TP_estimator} is evaluated exactly.
In contrast, empirical gradient (EG) estimators approximate the unknown gradient by replacing the activation probability $p_{\text{PSN}}\left(\mathbf{z}^{(l)}\right)$ with its empirical estimate $\hat{h}_{i}^{(l)}$ defined in Eq.~\eqref{eq:empirical_estimate}.
By definition, this yields an unbiased estimator of the activation probability, satisfying $\mathbb{E}\left[\hat{h}_{i}^{(l)}\right]=p_{\text{PSN}}\left(z_{i}^{(l)}\right)$.

The EG estimator is then constructed by leveraging the analytical form of the activation derivative, substituting $\hat{h}_{i}^{(l)}$ for the unknown true probability $p_{\text{PSN}}\left(z_{i}^{(l)}\right)$.
Concretely, assuming the derivative can be expressed as a function of the activation probability, the gradient with respect to the pre-activation in Eq.~\eqref{gz_PNN} is approximated as
\begin{equation}
    \frac{\partial L}{\partial \mathbf{z}^{(l)}} \approx \left(\left.\frac{\partial p_{\text{PSN}}\left(\mathbf{z}^{(l)}\right)}{\partial \mathbf{z}^{(l)}}\right|_{\hat{\mathbf{h}}^{(l)}}\right)^{\top} \ \frac{\partial L}{\partial \mathbf{h}^{(l)}},
\end{equation}
where the derivative $\partial p_{\text{PSN}}\left(\mathbf{z}^{(l)}\right)/\partial \mathbf{z}^{(l)}$ is evaluated by replacing the activation probability with its empirical estimate.

We stress that the construction of EG estimators relies on the autonomous representation of a functional form for the activation probability $p_{\text{PSN}}\left(z_{i}^{(l)}\right)$.
Specifically, the derivative of the activation probability needs to be expressible as a function of the probability itself, see App.~\ref{req_EG} for further details.

To illustrate this, here we present the result for PSNs whose activation probability follows a sigmoid-like form:
\begin{equation}
    p_{\text{PSN}}\left(z_{i}^{(l)}\right) = \sigma\left(z_{i}^{(l)}\right) = \frac{1}{1 + e^{-z_{i}^{(l)}}}
\end{equation}
Utilizing its autonomous representation, the derivative can be approximated using the empirical sample mean:
\begin{align}
    \frac{\partial p_{\text{PSN}}\left({z_{i}^{(l)}}\right)}{\partial z_{i}^{(l)}}
    &=
    p_{\text{PSN}}\left(z_{i}^{(l)}\right)\left(1 - p_{\text{PSN}}\left(z_{i}^{(l)}\right)\right) \nonumber\\
    &\approx
    \hat{h}_{i}^{(l)}\left(1 - \hat{h}_{i}^{(l)}\right).
\end{align}
Note that for sigmoid-like functions, this estimator becomes non-trivial only when $K \geq 2$, since for a single sample $\hat{h}_{i}^{(l)} \in \{0,1\}$ and the gradient vanishes.

Once the gradient with respect to the pre-activation has been obtained, parameter gradients follow standard backpropagation as in Eq.~\eqref{g_W_b}.
Thus, the EG estimator modifies only the local gradient computation at stochastic neurons, while leaving the overall backpropagation framework unchanged.
This approach embraces the stochasticity and discreteness of the sampling process during the backward computation, potentially acting as an implicit regularizer, similarly to stochastic regularization used to prevent overfitting.

\begin{figure}[!htbp]
\includegraphics[width=0.9\linewidth]{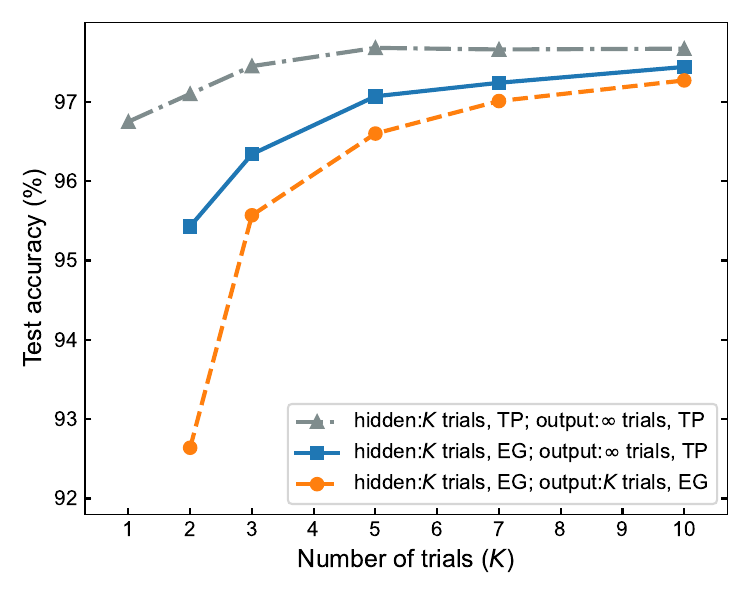}
\caption{\label{fig:empirical} 
Performance of the EG estimator under finite sampling constraints compared to the TP baseline.
The EG estimator applied to the hidden layer only and to both layers enables effective training under finite-sampling constraints ($K\leq10$).
Test accuracy quickly approaches TP upper bound as the trial count increases.
$K=1$ is omitted in all plots for the EG estimators in the hidden layer dues to vanishing empirical gradients.}
\end{figure}

Guided by the general requirement of autonomous representation, here we apply the EG estimator to SET neurons.
Fig.~\ref{fig:empirical} shows the test accuracy on MNIST digit classification obtained using the EG estimator, compared against a reference baseline.
We evaluate three training configurations: (i) a baseline applying the TP approach to both the hidden layer and output layer, (ii) applying the EG estimator only in the hidden layer while evaluating the output layer using the TP approach (infinite-trial limit), and (iii) applying the EG estimator in both the hidden and output layers by sampling the output to produce discrete class labels, since the softmax activation naturally defines a multinomial probability distribution.
The extension of the EG estimator to the softmax output layer is derived in App.~\ref{req_EG} and a smoothing technique to guarantee numerical stability under finite sampling at the output layer is detailed in Section~\ref{sec:output_trials}.
The configurations utilizing the EG estimator in the hidden layer are evaluated starting from $K=2$, since a single sample~($K=1$) causes the empirical gradient to vanish for the SET neuron's activation probability.
When comparing performance, the TP baseline acts as an upper bound.
When the EG estimator is applied only to the hidden layer, the network achieves high test accuracy that closely tracks the TP baseline with a relatively small performance gap. 
In contrast, applying the EG estimator to both layers introduces additional sampling noise, making the training more sensitive to the number of trials in the extremely low-sampling regime. 
However, as the number of trials increases, the performance of the fully empirical configuration improves rapidly and gradually converges toward the accuracy obtained in the hidden-only case.

\subsection{Straight-through estimators in backward pass}
\label{sec:STE}
Straight-through~(ST) estimators are a widely used heuristic for training neural networks with non-differentiable activation functions~\cite{bengio2013estimating,hubara2016binarized,yin2019understanding,wu2023estimator}.
The core idea of the ST estimator is to bypass both the stochastic sampling process and the detailed structure of the activation function during backpropagation, by replacing the true gradient with a surrogate that enables gradient-based optimization.

In the ST estimator, the non-differentiable operation in Eq.~\eqref{gz_PNN} is bypassed by defining a surrogate gradient in the backward pass:
\begin{equation}
    \frac{\partial \mathbf{h}^{(l)}}{\partial \mathbf{z}^{(l)}} \rightarrow \boldsymbol{\mathcal{S}}^{(l)},
\end{equation}
where $\boldsymbol{\mathcal{S}}^{(l)}$ denotes a surrogate derivative in layer $l$.
There is no unique choice of ST estimators.
A common choice is to simply set $\boldsymbol{\mathcal{S}}^{(l)}=\mathbb{I}$, the identity matrix. 

Under this replacement, the gradient with respect to the pre-activation, as seen in Eq.~\eqref{gz_PNN}, is given by
\begin{equation}
\frac{\partial L}{\partial \mathbf{z}^{(l)}} = \left(\boldsymbol{\mathcal{S}}^{(l)}\right)^{\top} \frac{\partial L}{\partial \mathbf{h}^{(l)}}.
\end{equation}
For the identity surrogate, this reduces to
\begin{equation}
    \frac{\partial L}{\partial \mathbf{z}^{(l)}} = \frac{\partial L}{\partial \mathbf{h}^{(l)}}.
\end{equation}

While this formulation applies directly to independent, element-wise activation in the hidden layers, using the ST estimator in the output layer requires additional consideration when the pre-activation is unknown.
For a sampled softmax output combined with CE loss, here the ST estimator is intuitively extended by replacing the true probability with the discrete sampled output.
This yields the pre-activation gradient:
\begin{equation}
    \frac{\partial L}{\partial \mathbf{z}^{(l)}} = \hat{\mathbf{p}} - \mathbf{y},
\end{equation}
where $\hat{\mathbf{p}}$ represents the discrete sampled output of the softmax layer, and $\mathbf{y}$ is the one-hot encoded label. 
The derivation justifying this extension of the ST estimator is detailed in App.~\ref{sec:ST_softmax}.

\begin{figure}[!htbp]
\includegraphics[width=0.9\linewidth]{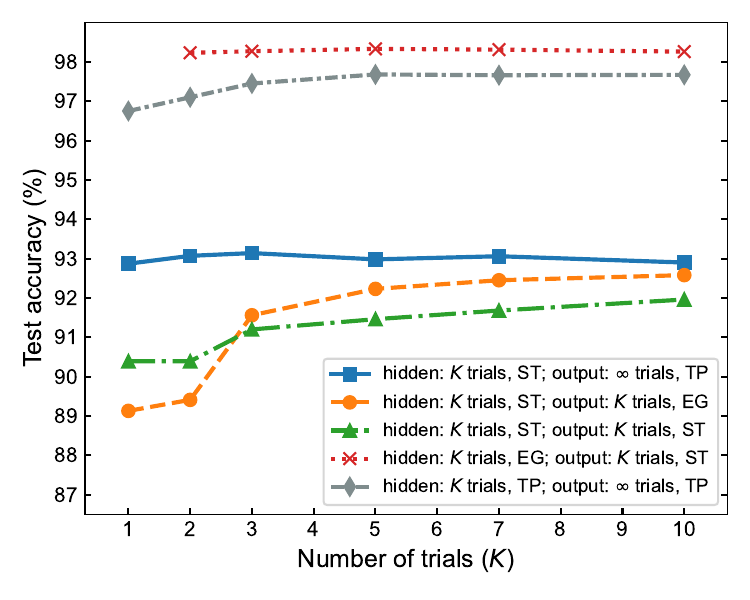}
\caption{\label{fig:ST} Performance of ST estimators under finite sampling.
Applying the identity ST estimator to the hidden layer limits the learning capacity, with test accuracy saturating below roughly $93\%$ across all evaluated output configurations.
In contrast, applying the EG estimator in the hidden layer while retaining a ST estimator in the output layer yields highly competitive performance ($>98\%$).
The configuration applying the TP approach in both hidden and output layers is a baseline.
Note that for $K=1$, the EG estimator at the output layer is enabled by introducing a smoothing parameter $\epsilon$ (detailed in Sec.~\ref{sec:output_trials}).
}
\end{figure}

Fig.~\ref{fig:ST} compares four estimator configurations obtained by combining ST and EG estimators, and the TP approach across the hidden and output layers.
When the identity ST estimator is used in the hidden layer with SET neurons featuring sigmoid-like activation probabilities, we evaluate three output-layer gradient estimators: TP, EG, and ST. 
In all three cases, the test accuracy improves monotonically with the number of trials.
However, the overall performance saturates at around $93\%$, even in the relatively large-trial regime.
In contrast, when the hidden layer is trained using the EG estimator and the output layer uses an ST surrogate, the network reaches approximately $98\%$ test accuracy with only a small number of trials.

\subsection{Infinite versus few trials in output layer}
\label{sec:output_trials}
In previous work on training stochastic neural networks~\cite{ma2025quantum,hubara2016binarized,rastegari2016xnor,zhou2016dorefa}, the output layer is typically evaluated via the true probability (i.e., infinite trials) approach.
By definition, the softmax function can also be treated as a multinomial distribution from which discrete class labels are drawn.
Beyond providing a fully stochastic forward pass, extending this sampling to the output layer offers a significant motivation: it holds the potential to be more energy-efficient.
In this case, each forward pass produces a discrete class label drawn from the softmax distribution, rather than an averaged probability vector.
However, introducing sampling at the output layer also directly affects the loss evaluation and gradient estimation, potentially impacting training stability and efficiency.

Specifically, finite sampling~($K<\infty$) at the softmax output introduces a numerical singularity when evaluating CE loss.
For any finite $K$, there exists a non-zero probability that the target class $i$ is not sampled.
This results in an empirical probability of $\hat{p}_{i}=0$, leading to undefined $\log(0)$ terms in the loss.
To address this issue in the finite-sampling regime, we introduce a sample smoothing technique inspired by label smoothing~\cite{szegedy2016rethinking}.
We construct a smoothed empirical probability $\hat{\mathbf{p}}^{s}$ by relaxing the raw empirical sample $\hat{\mathbf{p}}$:
\begin{equation}\label{eq:smoothed_p}
    \hat{\mathbf{p}}^{s} = (1-\epsilon)\hat{\mathbf{p}}+\frac{\epsilon}{C}\mathbf{1},
\end{equation}
where $\epsilon\in(0,1)$ is a smoothing factor and $C$ is the number of classes.
In the experiments, we empirically set $\epsilon=10^{-12}$.
This operation avoids the $\log(0)$ singularity in the cross-entropy loss for any finite $K$, thereby enabling effective gradient flow and stable training under finite-sampling constraints.

\begin{figure}[!htbp]
\includegraphics[width=0.9\linewidth]{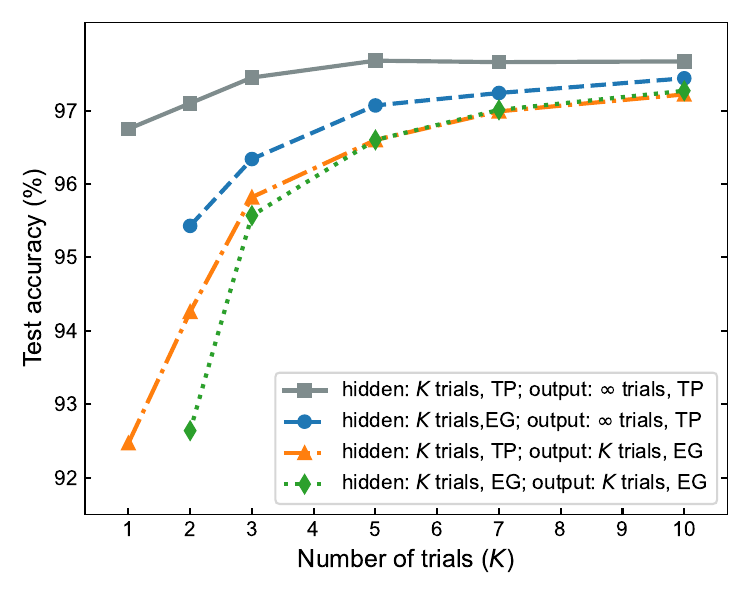}
\caption{\label{fig:infinite_vs_few} 
Output-layer performance under infinite and few trial regimes.
Finite output-layer sampling remains effective under strict trial-budget constraints ($K \leq 10$).
Test accuracy approaches the infinite-trial limit as the trial count increases.}
\end{figure}

Now we compare four training configurations that differ in how stochasticity is treated in the hidden and output layers.
In all cases, stochastic sampling is applied in the hidden layer across a range of trial budgets.
As reference baselines, we consider networks with output layers evaluated using infinite trials, trained using either the TP approach or the EG estimator in the hidden layer.
We then introduce output-layer sampling by drawing $K$ samples from the softmax distribution, matching the trial budget of the hidden layer, and perform training using the EG estimator.

Fig.~\ref{fig:infinite_vs_few} shows the results of these configurations.
When the output layer is evaluated in the infinite-trial limit, performance remains high and is relatively insensitive to the number of trials used in the hidden layer.
In contrast, introducing stochastic sampling at the output layer leads to a slight dependence on the trial budget.
As the number of trials increases, performance improves steadily.
In the intermediate-trial regime ($K \approx 5\,$--$\,10$), the gap between stochastic and deterministic output layers is substantially reduced.
In this regime, output-layer sampling becomes feasible though the EG estimator is used in the backward pass for both hidden layers and output layer.

\subsection{Softmax versus unnormalized activation in output layer}
\label{sec:soft_vs_linear}
While the combination of softmax activation and CE loss is a standard choice in digital neural networks~\cite{goodfellow2016deep}, it might be challenging in physical systems, as implementing a softmax operation together with sampling at the output layer would require a layer-wise global normalization across all output neurons.

To evaluate the trade-off, we compare two output-layer designs for stochastic PNNs: the softmax activation combined with the CE loss, and an alternative formulation based on linear output activations trained using the mean squared error (MSE) loss.

\begin{figure}[!htbp]
    \centering
    \includegraphics[width=0.9\linewidth]{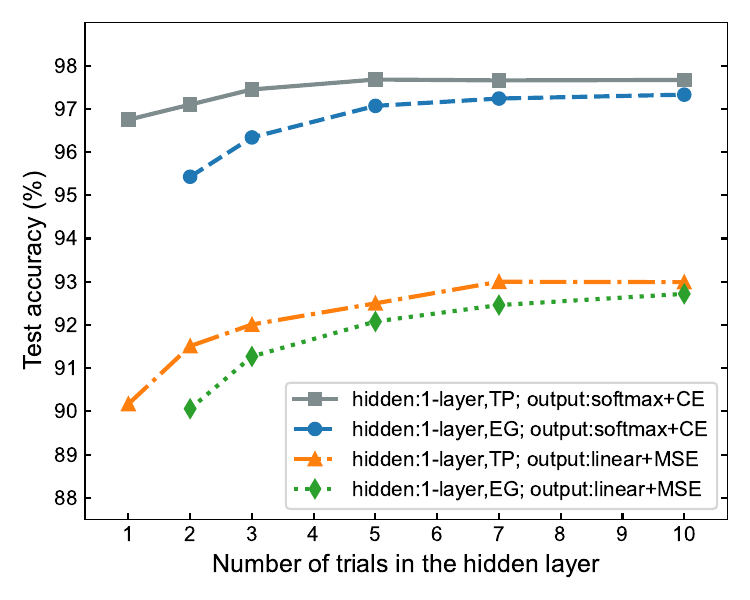}
    \centerline{(a) Single-hidden-layer networks}
    \includegraphics[width=0.9\linewidth]{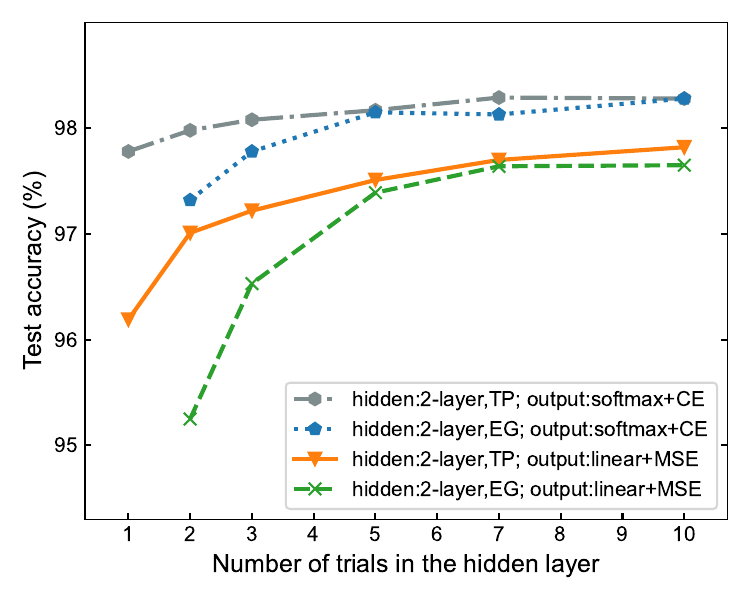}
    \centerline{(b) Two-hidden-layer networks}
    \caption{Comparison of test accuracy using softmax and unnormalized linear activations in the output layer for models with (a) a single hidden layer and (b) two hidden layers.
    While the linear and MSE formulation underperforms in shallow networks, increasing the network depth to 2 hidden layers substantially improves its accuracy, making it comparable to the 1-layer softmax and CE configuration.}
    \label{fig:softmax_vs_linear}
\end{figure}

Fig.~\ref{fig:softmax_vs_linear} compares the performance of these two configurations with different settings.
To ensure a fair comparison, the output layer in all evaluated models is set to the infinite-trial limit and updated using the TP approach.
For shallow networks comprising a single hidden layer, the softmax-based models demonstrate strong performance, reaching roughly $97.5\%$ when applying either the TP approach or the EG estimator to the hidden layer.
In contrast, the linear and MSE formulation exhibits significantly lower performance in the same architecture, saturating below approximately $93\%$.
This gap may be attributed to the lack of nonlinearity at the output layer.
With the addition of a second hidden layer, the performance of the linear and MSE formulation improves substantially, reaching test accuracies comparable to the single-hidden-layer softmax baseline. Furthermore, applying the 2-layer architecture to the softmax and CE configuration yields the highest overall performance, exceeding 98\%.

\section{Conclusions}
In this paper, we investigated the training of PNNs in the extreme limit of energy efficiency, where information is carried by discrete physical quanta and intrinsic noise becomes unavoidable.
Instead of treating device stochasticity as perturbative errors to be suppressed, we presented a training framework that explicitly incorporates PSN behavior.
By modeling physically-motivated neurons based on single-electron transistors and single-photon sources, we presented alternative pathways to the realization of stochastic PNNs.
The true single-photon PSN, in particular, offers a potential pathway to the realization of fully quantum stochastic PNNs, including the possibility of any quantum advantage they might offer.

A key challenge is that the exact activation probabilities or pre-activation values are generally inaccessible during training.
To address this issue, we introduced the EG estimator, which enables gradient-based training using a limited number of discrete output samples, unlike the TP approach~\cite{ma2025quantum} that relies on an infinite-trial limit.
Numerical experiments on the MNIST dataset demonstrated that reliable training remains possible in a very few-sample regime even when the output layer is discrete. 
Furthermore, by exploring various training configurations, such as combining the EG estimator in the hidden layers with a ST estimator, or an unnormalized linear activation at the output, we showed that effective training can be maintained while minimizing experimental sampling cost.

This work illustrates a pathway toward bridging the gap between neural network training algorithms and physically realizable computation.
By incorporating the properties of physical devices into the training procedure, stochastic PNNs may enable energy-efficient information processing beyond the limits of conventional neural networks.
Although challenges remain in scaling these architectures and implementing fully in-situ training, the methods developed in this work provide a practical framework for learning in stochastic PNNs.

\begin{acknowledgments}
T.D. thanks Daoyi Dong for helpful discussions.
\end{acknowledgments}

\appendix

\section{Analysis of true single-photon stochastic neuron}\label{append_ps_neuron}
As described in Sec.~\ref{sec:TSP_neuron} of the main text, we consider a system composed of two bosonic modes, denoted by $a$ and $b$.
These modes interact via controllable beamsplitter Hamiltonian in Eq.\eqref{eq:Hamiltonian} of the main text,
\begin{equation}
    H = \alpha(t) a b^\dagger +\alpha^* (t)a^\dagger b,
\end{equation}
where $\alpha(t)$ is the time-dependent coupling strength.
This Hamiltonian describes general SU(2) mixing between two modes, which we can realize with various engineered systems. 
In the context of the optomechanical implementation discussed in the main text, $a$ refers to an optical cavity mode and $b$ refers to a mechanical oscillator mode.

The joint system Hilbert space for the $b$ mode and the $a$ mode is $\mathcal{H}_b\otimes\mathcal{H}_a$. Utilizing the SLH framework \cite{Combes_2017}, we consider both modes to be coupled to fields $\mathcal{F}$ described by a continuous-mode bosonic symmetric Fock space of traveling wave packets with temporal envelope $\xi(t)$, normalised such that $\int_0^\infty |\xi(t)|^2dt=1$. 
Eq.~\eqref{eq:single_photon_state} of the main text defines a single-photon state as a coherent superposition over frequency modes $\omega$ of a vacuum field excitation
\begin{equation}
   |1_\xi\rangle=\int d\omega\tilde{\xi}(\omega)a_{\omega}^\dagger|0\rangle,
\end{equation}
where $|0\rangle\in\mathcal{F}$ is the incoming field vacuum state and $a_\omega$ is an annihilation operator for the field such that $[a_\omega,a_{\omega'}^\dagger]=\delta(\omega-\omega')$ and $a_\omega|0\rangle=0$ for all $\omega$. 
The setting for the system and the input field is $\mathcal{H}_b\otimes\mathcal{H}_a\otimes\mathcal{F}$.

In the FSME, the system is extended to couple to a photon in an external field. 
The time evolution of system operators is given by 
\begin{equation}
X(t)\equiv j_t(X)=U^\dagger(t) \left (X\otimes\mathbb{I}_f\right )U(t),
\end{equation}
where $X$ is a system operator and $\mathbb{I}_f$ is the identity operator acting on the fields and $U(t)$ acts nontrivially on the combined system and field modes.

For the b mode, the fact that the system contains only a single excitation, and that there is no explicit coupling from the $b$ mode to the single-photon field means that the dynamics $b(t)$, $b^{\dagger}(t)$ and $b^\dagger b(t)$ have no field operators; they act trivially on the field Hilbert space. 
As a result, the system-field time evolution can be thought of in terms of just system dynamics; they are pure system observables. 
This results in the ability to factorize observables. 
This factorization does not hold for the $a$ mode, whose direct coupling to the external field introduces explicit field operators into its Heisenberg evolution.

The inner products of time-evolved system operators, $\varpi^{ij}_t(X)$, are
\begin{equation}
    \varpi^{ij}_t(X)=\langle \eta~\phi_i|j_t(X)|\eta~\phi_j\rangle,
\end{equation} 
where we write $|\eta~\phi_k\rangle=|\eta\rangle\otimes|\phi_k\rangle\in\mathcal{H}_a\otimes\mathcal{H}_b\otimes\mathcal{F}_{a}$, for the initial state $|\eta\rangle\in\mathcal{H}_a\otimes\mathcal{H}_b$ of the system and with $|\phi_k\rangle\in\mathcal{F}_{a}$ being $|0\rangle$ for $k=0$ and $|1_\xi\rangle$ for $k=1$. 
Equivalently, $\varpi^{ij}_t(X)$ are associated with trace class operators $\rho^{ij}(t)$ such that, as defined in \cite{gough2012quantum},
\begin{equation}
    \mathrm{Tr}\left(\rho^{ij}(t)^\dagger j_t(X)\right)=\varpi^{ij}_t(X).
\end{equation}

By re-writing Eq.~\eqref{eq:varphi_as_expectation_1} as a trace inner product, we can identify these operators $\rho^{ij}(t)^\dagger$  as $|\eta~\phi_j\rangle\langle\eta~\phi_i|$
\begin{equation}
    \varpi^{ij}_t(b^\dagger b)=\mathrm{Tr}\Big(\underbrace{|\eta~\phi_j\rangle\langle \eta~\phi_i|}_{=\rho^{ij}(t)^\dagger}j_t(b^\dagger b)\Big),
\end{equation}
\noindent which is such that
\begin{equation}
    \rho^{ij}(t)^\dagger=(|\eta~\phi_j\rangle\langle \eta~\phi_i|)^\dagger=|\eta~\phi_i\rangle\langle \eta~\phi_j|=\rho^{ji}(t).
\end{equation}
For $\varpi^{11}_t(b^\dagger b)$, which is the expected value of $b^{\dagger}b$ given a single-photon excitation in the field coupled to mode $a$,
\begin{equation}
\begin{split}
    \varpi^{11}_t(b^\dagger b)&=\mathrm{Tr}\{j_t(b^\dagger b) \rho^{11}(t)\}\\
    &=\mathrm{Tr}\{U^\dagger(t)\left(b^\dagger b\otimes\mathbb{I}_f\right)U(t) (\rho^{01})^\dagger(\rho^{10})^\dagger\}\\
    &=\mathrm{Tr}\{U^\dagger(t)\left(b^\dagger\otimes\mathbb{I}_f\right)\left(b\otimes\mathbb{I}_f\right)U(t) (\rho^{01})^\dagger(\rho^{10})^\dagger\}.\\
\end{split}
\end{equation}
Because there is no explicit coupling between the single photon source and the $b$ mode, the dynamics $b(t),~b^\dagger(t)$ and $b^\dagger b(t)$ are not associated with field operators, hence we make the ansatz that trace factorization occurs such that
\begin{align}
    \varpi^{11}_t(b^\dagger b)&=\mathrm{Tr}\left\{U^\dagger(t)\left(b^\dagger\otimes\mathbb{I}_f\right)\left(b\otimes\mathbb{I}_f\right)U(t) (\rho^{01})^\dagger(\rho^{10})^\dagger\right\}\nonumber\\
    &=\mathrm{Tr}\left\{(\rho^{10})^\dagger U^\dagger(t)\left(b^\dagger\otimes\mathbb{I}_f\right)U(t)\right\}\nonumber\\
    & \quad\,\,\,\, \times\mathrm{Tr}\left\{U^\dagger (t)\left(b\otimes\mathbb{I}_f\right)U(t) (\rho^{01})^\dagger\right\}\nonumber\\
    &=\mathrm{Tr}\{(\rho^{10})^\dagger j_t(b^\dagger)\}\mathrm{Tr}\{j_t(b) (\rho^{01})^\dagger\}\nonumber\\
    &=\varpi^{10}(b^\dagger)\varpi^{01}(b)\nonumber\\
    &=\varpi^{10}(b^\dagger)\varpi^{10}(b^\dagger)^*,
\end{align}
where the final line utilizes $\varpi^{ij}(X^\dagger)^*=\varpi^{ji}(X)$. Hence it suffices to calculate $\varpi^{10}_t(b^\dagger)$ and multiply it by its complex conjugate to obtain $\varpi^{11}_t(b^\dagger b)$.

The direct application of the Cauchy-Scwartz inequality on $\varpi^{10}(b^\dagger)\varpi^{10}(b^\dagger)^*=|\varpi^{10}(b^\dagger)|^2$ provides confirmation that the right-hand-side is bounded by 1, as we would expect given the system is driven by a single excitation. 
As a trace inner product $\langle A,B\rangle_{\mathrm{Tr}}=\mathrm{Tr}(B^\dagger A)$ between $j_t(b^\dagger)$ and $\rho^{01}(t)$, we have
\begin{equation}
    \varpi^{10}_t(b^\dagger)=\mathrm{Tr}\left(j_t(b^\dagger)\rho^{01}(t)\right)=\langle \rho^{01}(t),j_t(b^\dagger)\rangle_{\mathrm{Tr}}
\end{equation}
such that
\begin{align}
   \varpi^{10}(b^\dagger)\varpi^{10}(b^\dagger)^*
   &\leq  \langle \rho^{01}(t),\rho^{10}(t)\rangle_{\mathrm{Tr}}\langle j_t(b^\dagger),j_t(b^\dagger)\rangle_{\mathrm{Tr}} \nonumber \\
   &=\mathrm{Tr}\left( |\eta~1_\xi\rangle\langle\eta~0|\eta~0\rangle\langle\eta~1_\xi| \right) \mathrm{Tr}\left( j_t(b)j_t(b^\dagger) \right) \nonumber \\
   &=\mathrm{Tr}\left( |\eta~1_\xi\rangle\langle\eta~1_\xi| \right)\mathrm{Tr}\left( j_t(b b^\dagger)\right) \nonumber \\
   &=\varpi^{11}_t(\mathbb{I})\mathrm{Tr}\left( j_t(b b^\dagger)\right)
\end{align}
We have $\varpi^{11}_t(\mathbb{I})=1$ and for a single excitation we have the matrix representation
\begin{equation}
    b b^\dagger=\begin{bmatrix}
        1 & 0\\
        0 & 0
    \end{bmatrix},
\end{equation}
hence $\mathrm{Tr}\left( j_t(b b^\dagger)\right)=1$ and so $\varpi^{10}(b^\dagger)\varpi^{10}(b^\dagger)^*\leq 1$, as claimed.

The coupled dynamics of inner products of system operators evolve according to Eq.~\eqref{eq:monent_omega} of the main text, which we repeat here for convenience.
The equation for $\varpi^{10}_t(X)$ that follows from Eq.~\eqref{eq:monent_omega} is the equation 
\begin{equation}
    \dfrac{d}{dt}\!
    \begin{bmatrix}
        \varpi^{10}_t(a^\dagger)\\
        \varpi^{10}_t(b^\dagger)
    \end{bmatrix}
    \!\!=\!\!
    \begin{bmatrix}
        \varpi^{10}_t(\mathcal{L}a^\dagger)+\varpi^{00}_t([a^\dagger,\sqrt{\kappa} a])\xi(t)\\
        \varpi^{10}_t(\mathcal{L}b^\dagger)
    \end{bmatrix}\!.
\end{equation}
Note the second term in the second component is zero since there is no photon input into the $b$ mode. Assuming real and constant $\alpha$ and truncating expressions involving more than two ladder operators, the Lindbladians are $\mathcal{L} a^\dagger =i\alpha(t) b^\dagger-\dfrac{\kappa}{2}a^\dagger$ and $\mathcal{L} b^\dagger=i\alpha^*(t) a^\dagger-\dfrac{\gamma}{2} b^\dagger$.

Now we consider a normalised exponentially decaying single photon source on the cavity mode only
\begin{equation}
    \xi(t)=\begin{cases}
        \sqrt{\zeta}e^{-\zeta t/2}, & t\geq 0\\
        0, & t<0
    \end{cases}~.
\end{equation}

Since $ \varpi^{00}_t([a^\dagger,\sqrt{\kappa} a])\xi(t)=-\sqrt{\kappa}\xi(t)$, combining the above we arrive at the affine equation,
\begin{equation}\label{eq:matrixEquation}
    \dfrac{d}{dt} 
    \begin{bmatrix}
        \varpi^{10}_t(a^\dagger)\\
        \varpi^{10}_t(b^\dagger)
    \end{bmatrix}
    =
    \begin{bmatrix}
        -\kappa/2 & i\alpha^*(t)\\
        i\alpha(t) & -\gamma/2
    \end{bmatrix}
    \begin{bmatrix}
        \varpi^{10}_t(a^\dagger)\\
        \varpi^{10}_t(b^\dagger)
    \end{bmatrix}
    +
    \begin{bmatrix}
        -\sqrt{\kappa}\xi^*(t)\\
         0
    \end{bmatrix}.
\end{equation}
The solution to Eq.~\eqref{eq:matrixEquation} can be expressed in terms of a Green's matrix $\mathbf{G}$ as
\begin{equation}
    \begin{bmatrix}
        \varpi^{10}_t(a^\dagger)\\
        \varpi^{10}_t(b^\dagger)
    \end{bmatrix} 
    =\int_0^t \mathbf{G}(t,t') 
    \begin{bmatrix}
        -\sqrt{\kappa}\xi(t')\\
         0
    \end{bmatrix}
    dt'.
\end{equation}

\noindent If we take $\alpha(t)$ to be a rectangular pulse beginning at $t=0$ applied for a time period $T$, we would have a real and constant $\alpha(t)=\alpha$ up to $t=T$. The Green's function under this assumption for $t< T$ reduces to
\begin{equation}
    \mathbf{G}(t,t')=e^{\mathbf{M} (t-t')},
\end{equation}

\noindent where
\begin{equation}
   \mathbf{M}= 
   \begin{bmatrix}
        -\kappa/2 & i\alpha\\
        i\alpha & -\gamma/2
    \end{bmatrix}.
\end{equation}
Then Eq.~\eqref{eq:matrixEquation} can be solved by diagonalizing. 
We obtain 
\begin{equation}
\begin{split}
\varpi^{10}_t(b^\dagger)
&=
\frac{2 i \alpha \sqrt{{\zeta}}\sqrt{\kappa}}{\Delta}
\Bigg[
\frac{
4 - 4  e^{\frac{t}{4}\left(\Delta - u\right)}
}{
e^{\frac{t}{2}{\zeta}}
\left(\Delta - u\right)
}
\\[4pt]
&\qquad
+
\frac{
4 e^{-\frac{t}{2}{\zeta}}
- 4 e^{-\frac{t}{4}(\gamma + \Delta + \kappa)}
}{
\Delta + u
}
\Bigg],
\end{split}
\end{equation}
where
\begin{subequations}
    \begin{align}
        \Delta &= \sqrt{(\gamma - \kappa)^2 - 16\alpha^2},\\
        u &= \gamma + \kappa -2\zeta.
    \end{align}
\end{subequations}

\noindent Due to the factorization $\varpi^{10}(b^\dagger)\varpi^{10}(b^\dagger)^*$, we need only multiply $\varpi^{10}_t(b^\dagger)$ by its conjugate to obtain the expected occupation of the $b$ mode as in Eq.~\eqref{eq:ps_activation} of the main text,
\begin{equation}
\!\!\langle b^\dagger b\rangle \!=\! \frac{64\alpha^{2}{\zeta} \kappa}{\Delta^2}
\!\Bigg[
\frac{1 - e^{\frac{t}{4}\left(\Delta - u\right)}}{\Delta - u} 
+
\frac{1 - e^{\frac{t}{4}\left(\Delta + u \right)}}{\Delta + u}
\Bigg]^{2}
\!\!e^{-t {\zeta}}.
\end{equation}
This goes to zero as $\alpha\to 0$ as needed, and we have $\varpi^{11}_t(b^\dagger b)=\varpi^{11}_t(b^\dagger b)^*$, also as expected.

\section{Requirements for EG estimators}\label{req_EG} 
To apply the EG estimators introduced in Sec.~\ref{sec:EG} without explicitly evaluating the pre-activation, we require that the derivative of the activation probability $p(z)$ need to be written as a function of $p(z)$ itself; i.e.,
\begin{equation}
\frac{\partial p(z)}{\partial z} = g(p(z))
\end{equation}
for some well-defined function $g: \mathbb{R}\to\mathbb{R}$.

\subsection{Necessary and sufficient condition for autonomous representation}
We now state a simple characterization of when such an autonomous representation exists.
Let $I \subseteq \mathbb{R}$ be an interval and let $p: I \to \mathbb{R}$ be a continuously differentiable function.
There exists a function $g: p(I) \to \mathbb{R}$ such that the derivative of $p$ can be expressed as a function of $p$ itself, i.e.,
\begin{equation}
p'(z) = g\left(p(z)\right), \quad \forall~z \in I
\end{equation}
if and only if the following condition holds for all $z_1, z_2 \in I$:
\begin{equation}
p(z_1) = p(z_2) \implies p'(z_1) = p'(z_2) \label{eq:condition}
\end{equation}
Furthermore, if condition \eqref{eq:condition} is satisfied, the function $g$ is uniquely determined on the image $p(I)$ by:
\begin{equation}
g(y) = p'(z) \quad \text{for any } z \in I \text{ such that } p(z) = y
\end{equation}
In this case, $g$ is said to be well-defined.

\subsection{Strict monotonicity as a sufficient condition}
For PSNs whose activation probability $p(z)$ is a strict monotonic function of the pre-activation $z$, condition~\eqref{eq:condition} is automatically satisfied.
This monotonicity guarantees that $p(z)$ is a one-to-one mapping.

In this scenario, any given probability $y=p(z)$ corresponds to a unique pre-activation $z$, allowing us to define the inverse function $z=p^{-1}(y)$.
Consequently, the autonomous derivative function $g$ can be directly constructed by evaluating the derivative $p'$ at this inverse mapping:
\begin{equation}
    g(y) = p'\bigl(p^{-1}(y)\bigr), \quad \forall~y \in p(I).
\end{equation}
By substituting the unique $z = p^{-1}(y)$ into the relation $p'(z) = g(p(z))$, we can recover $p'(p^{-1}(y))=g(y)$.
This provides a straightforward, closed-form expression for deriving the EG estimator whenever the inverse of the PSN's activation function is known. 

For the sigmoid activation $p(z)=\sigma(z)=1/(1+e^{-z})$, the inverse is the logit function $z=\ln \left(y/(1-y)\right)$. Substituting this into $p^{'}(z)=p(z)\left(1-p(z)\right)$ yields $g(y)=y(1-y)$, which is independent of $z$.

\subsection{EG estimator for softmax activation}
Here we show that the EG estimator can be extended to the softmax activation (i.e., its Jacobian can be expressed solely in terms of the softmax outputs), and derive the corresponding form.

Let $\mathbf{z} = (z_1, z_2, \ldots, z_n)^\top \in \mathbb{R}^n$ be the input vector to the softmax function, and let $\mathbf{p} = (p_1, p_2, \ldots, p_n)^\top$ be the output vector defined by
\begin{equation}
p_i = \text{softmax}(z_{i}) = \frac{e^{z_i}}{\sum_{j=1}^n e^{z_j}}, \quad i = 1, 2, \ldots, n.
\end{equation}

The Jacobian matrix $\mathbf{J} \in \mathbb{R}^{n \times n}$ of the softmax function has entries given by
\begin{equation}
\mathbf{J} _{ij} = \frac{\partial p_i}{\partial z_j} = p_i (\delta_{ij} - p_j),
\end{equation}
where $\delta_{ij}$ is the Kronecker delta. In matrix form, the Jacobian can be expressed as
\begin{equation}
\mathbf{J} = \text{diag}(\mathbf{p}) - \mathbf{p} \mathbf{p}^\top,
\end{equation}
where $\text{diag}(\mathbf{p})$ is a diagonal matrix with the elements of $\mathbf{p}$ on the diagonal.
To construct the EG estimator for the softmax activation, we replace the true probabilities $\mathbf{p}$ with their empirical estimates $\hat{\mathbf{p}}$ obtained from finite sampling.
Thus, the empirical Jacobian matrix $\hat{\mathbf{J}}$ is given by
\begin{equation}\label{eq:empirical_Jacobian}
\hat{\mathbf{J}} = \text{diag}(\hat{\mathbf{p}}) - \hat{\mathbf{p}} \hat{\mathbf{p}}^\top.
\end{equation}

Note that in the single-trial case, the pure empirical Jacobian $\hat{\mathbf{J}}$ becomes degenerate.
In this situation, the empirical probability $\hat{\mathbf{p}}$ collapses to a one-hot vector.
Substituting this one-hot vector into Eq.~\eqref{eq:empirical_Jacobian} yields $\text{diag}(\hat{\mathbf{p}}) = \hat{\mathbf{p}} \hat{\mathbf{p}}^\top$, which reduces $\hat{\mathbf{J}}$ to a zero matrix.
This degeneracy is mitigated by the sample smoothing technique introduced in Eq.~\eqref{eq:smoothed_p}, resulting in $\hat{\mathbf{p}}^{s}$.
The smoothed empirical Jacobian $\hat{\mathbf{J}}^{s}$ is given by
\begin{equation}
    \hat{\mathbf{J}}^{s} = \text{diag}(\hat{\mathbf{p}}^{s}) - \hat{\mathbf{p}}^{s} \hat{\mathbf{p}}^{s\top}.
\end{equation}
Because $\epsilon>0$ ensures that all elements of $\hat{\mathbf{p}}^s$ are within $(0, 1)$, we have $\hat{p}_i^s > (\hat{p}_i^s)^2$ for all $i$.
Consequently, the diagonal entries of $\hat{\mathbf{J}}^s$ remain strictly positive, ensuring a non-vanishing Jacobian in the single-trial case.

\section{ST estimator for softmax activation and cross-entropy loss}
\label{sec:ST_softmax}
This appendix details the surrogate gradient for the CE loss when the output-layer softmax is subject to finite sampling and the backward pass utilizes an ST estimator.

In the forward pass, the probability distribution $\mathbf{p}$ generated by the softmax is subject to a discrete sampling  process.
The resulting empirical estimate $\hat{\mathbf{p}}$ is given by
\begin{equation}
    \hat{\mathbf{p}} \sim \text{Multinomial}\left(K,\mathbf{p}\right),
\end{equation}
where $K$ is the number of trials.

Let $\mathbf{y}\in\mathbb{R}^{n}$ be a one-hot encoded target label. Then the standard CE loss evaluated on $\hat{\mathbf{p}}$ is defined as
\begin{equation}
    L = -\sum_{k}y_{k}\log \hat{p}_{k}.
\end{equation}
Thus, the derivative of the loss with respect to the empirical sampled output is $\partial L / \partial \hat{p}_{k}=- y_{k} / \hat{p}_{k}$. 
To backpropagate the gradient to the pre-activation $z_{i}$ at the output layer, we apply the chain rule,
\begin{equation}\label{eq:STE_scalar}
    \frac{\partial L}{\partial z_{i}} = \sum_{j}\sum_{k} \frac{\partial L}{\partial \hat{p}_{k}} \frac{\partial \hat{p}_{k}}{\partial p_{j}}\frac{\partial p_{j}}{\partial z_{i}}.
\end{equation}

To resolve the non-differentiable sample step, here we introduce two approximations to the backward pass. 
First, suppose that the sampling operation is bypassed by setting the surrogate gradient of the sample with respect to the probability as the identity matrix, i.e., $\partial \hat{p}_{k} / \partial p_{j}=\delta_{jk}$.
Then, since we assume that $\mathbf{p}$ is not directly observable in the forward pass, the unknown $p_{j}$ is replaced by its estimate $\hat{p}_{j}$,
\begin{equation}
    \frac{\partial p_{_{j}}}{\partial z_{i}} = p_{j}\left(\delta_{ij}-p_{i}\right) \approx \hat{p}_{j}\left(\delta_{ij}-\hat{p}_{i}\right).
\end{equation}
Applying these approximations to the Eq.~\eqref{eq:STE_scalar} provides the surrogate gradient:
\begin{align}
    \frac{\partial L}{\partial z_{i}} &\approx \sum_{j}  -\frac{y_{j}}{\hat{p}{j}} \cdot 1 \cdot \hat{p}_{j}(\delta_{ij} - \hat{p}_{i}) \nonumber \\
    &= -y_{i}(1-\hat{p}_{i}) + \sum_{j\neq i} y_{j}\hat{p}_{i} \nonumber \\
    &= -y_{i} + \sum_{j} y_{j} \hat{p}_{i} \nonumber \\
    &= \hat{p}_{i} - y_{i},
\end{align}
which in vector form is 
\begin{equation}
    \frac{\partial L}{\partial \mathbf{z}} \approx \mathbf{\hat{p}} - \mathbf{y}.
\end{equation}
Note that here we suppose that the cancellation $\partial \hat{p}_{j} / \partial \hat{p}_{j}=1$ even when a class is not sampled among the $K$ trials.
\vspace*{\fill}


\bibliography{apssamp}

\end{document}